\documentclass[a4paper,fleqn,usenatbib]{mnras}

\usepackage{newtxtext,newtxmath}

\usepackage[T1]{fontenc}
\usepackage{ae,aecompl}

\usepackage{graphicx}	
\usepackage{amsmath}	
\usepackage{amssymb}	

\title[Exotopography]{Finding Mountains with Molehills:\\ The Detectability of Exotopography}

\author[McTier \& Kipping]{
Moiya A.S. McTier$^{1}$,
David M. Kipping$^{1}$
\\
$^{1}$Department of Astronomy, Columbia University, 550 W 120th St., New York, NY 10027
}

\date{Accepted 2018 January 15. Received 2018 January 14; in original form 2017 August 8.}

\pubyear{2017}

\begin{document}
\label{firstpage}
\pagerange{\pageref{firstpage}--\pageref{lastpage}}
\maketitle

\begin{abstract}
Mountain ranges, volcanoes, trenches, and craters are common on rocky bodies throughout the Solar System, and we might we expect the same for rocky exoplanets. With ever larger telescopes under design and a growing need to not just detect planets but also to characterize them, it is timely to consider whether there is any prospect of remotely detecting exoplanet topography in the coming decades. To test this, we devised a novel yet simple approach to detect and quantify topographical features on the surfaces of exoplanets using transit light curves. If a planet rotates as it transits its parent star, its changing silhouette yields a time-varying transit depth, which can be observed as an apparent and anomalous increase in the photometric scatter. Using elevation data for several rocky bodies in our solar system, we quantify each world's surface integrated relief with a ``bumpiness'' factor, and calculate the corresponding photometric scatter expected during a transit. Here we describe the kinds of observations that would be necessary to detect topography in the ideal case of Mars transiting a nearby white dwarf star. If such systems have a conservative occurrence rate of 10\%, we estimate that the upcoming Colossus or OWL telescopes would be able to detect topography with $<$20 hours of observing time, which corresponds to $\sim$400 transits with a duration of 2 minutes and orbital period of $\sim$10 hours.
\end{abstract}




\section{Introduction}
\label{sec:intro}
The discovery of thousands of exoplanets in the last two decades has revealed that planets are very common around other stars \citep{dressing:2015}, but the occurrence rate of planets that truly resemble the Earth remains unknown. Finding that occurrence rate required a shift in focus from planet discovery to planet characterization. Along these lines, the search for Earth-sized planets located within the surface liquid water zones of their stars has already yielded several discoveries (e.g. \citet{quintana:2014, jenkins:2015, gillon:2017}).

A major challenge for modern astronomy is to advance our understanding beyond whether these planets have the same size and insolation as the Earth, to truly understand their characteristics and ultimately the uniqueness of our own home. Such an endeavour requires not only advances in our instrumentation, but also a paradigm shift in the way we model our observations, from point-masses described by a few parameters, to rich, textured globes with diverse environments and qualities.

An obvious step in characterizing an exoplanet is studying its atmosphere via methods such as transmission spectroscopy \citep{seager:2000} and indeed considerable research energy is being deployed in this area. It has been suggested that exoplanetary characterization need not be limited to atmospheric inference, though. For example, \citet{robinson:2010} argue that ocean glint would also produce an observable signature. Other literature examples include the observational signatures expected due to planetary  oblateness \citep{seager:2002, carter:2010}, circumplanetary rings \citep{arnold:2006}, exomoons \citep{sartoretti:1999, kipping:2009, kipping:2015}, industrial pollution \citep{lin:2014}, night lights in alien cities \citep{loeb:2012}, and plant pigments \citep{berdyugina:2016}. Although many of these effects are not immediately detectable with current facilities, these works provide the first estimates of what the limits of exoplanet characterization are and what kind of observatories the community requires to achieve detailed remote sensing of exoplanets.

Along these lines, a natural question to ask is whether surface features of exoplanets, namely mountains, volcanoes, trenches, and craters, could also impose an observable signature in astronomical data. In this paper, we explore the possibility of constraining a planet's surface integrated relief -- which we call ``bumpiness" -- by considering the scatter produced in a light curve as a transiting planet rotates and its silhouette changes. Such a constraint can be used to infer other planet characteristics, discussed further in \S~\ref{sec:discussion}.

Advancing exoplanet characterization beyond atmospheric retrieval is undoubtedly challenging, but ultimately the field of exoplanetary science is maturing to a long-game mindset where major breakthroughs will require planning for future missions. Groundbreaking telescopes like the Extremely Large Telescope (ELT), the Overwhelmingly Large Telescope (OWL), and Colossus \citep{kuhn:2014} would have the ability to provide superior photometry to current state-of-the-art. This opens the door to pursuing more challenging measurements, and the time to imagine what those measurements could be is now.

The goal of this paper is to provide a first estimate of the detectability of exotopgraphy by considering the increased photometric scatter in the light curve caused by topographical surface features. Whilst the idea of detecting such features is surely impractical presently, the planned very large telescopes of the coming decades are expected to lead to dramatic improvements in sensitivity and thus exotopgraphy may be feasible in the next few decades. 

In \S~\ref{sec:concept}, we outline the thesis of our detection method in more detail. We describe the process we used to derive our relationship between scatter and bumpiness in \S~\ref{sec:model}. In \S~\ref{sec:feasibility}, we discuss the feasibility of our method and describe the kinds of observations that would be necessary to detect topography given certain ideal conditions. In \S~\ref{sec:discussion}, we discuss what we can learn from the quantification of a planet's bumpiness and what can be done in the future to expand upon and improve this work.

\section{Observable Signatures of Topography}
\label{sec:concept}
One can conceive of several ways in which exoplanet topography may reveal itself. At the simplest level, direct imaging of an exoplanet could directly detect such structure, although the angular resolution required, of order ${\sim}10$\,km$/10$\,pc $\sim$ 10 nanoarcseconds, is far beyond current or planned capabilities. \citet{fujii:2014} propose using the albedo profiles of directly imaged exoplanets to study surface composition, which might be used to infer the presence of some topography.

Another possibility is to detect exomoons that transit a mountainous, luminous planet (from either thermal emission or reflected light). This would provide a raster-scan of the planet's luminosity profile, and is similar to the method described in \citet{cabrera:2005}. In order to disentangle from the star, nulling of the star would greatly help, although in principle sufficiently precise photometry could work \citep{forgan:2017}.

Imaging aside, exoplanet detection methods that seek the gravitational influence of the planet, such as radial velocities, microlensing, astrometry, and timing, will provide no observable signature of exotopography. This leaves the transit method as the only other option for seeking such signatures.

The transit method directly measures the sky-projected area of a planet's silhouette relative to that of a star, under the assumption that the planet is not luminous itself. A transit's depth, $\delta$, is given by  

\begin{equation}\label{depth}
\delta = \frac{A_P}{\pi R_*^2},
\end{equation}

where $A_P$ is the sky projected area of the planet and $R_*$ is the radius of the planet's host star. This fact implies that there is indeed some potential for transits to reveal surface features, since the planet's silhouette is certainly distorted from a circular profile due to the presence of topography. Whilst numerous strategies may exist for exploiting this point, we consider a scatter based approach for the sake of simplicity and presenting a first-order evaluation of the detectability of exotopography.

Before describing a data-driven model along these lines, we first outline a simple toy model as a pedagogical example.

Consider an Earth-sized planet that is perfectly spherical except for a block of material as tall as Everest, as long as the Himalayan mountain range running from north to south, and wide enough to span $1^{\circ}$ in longitude. Now assume that the planet completes half of one rotation as it transits its parent star from our point of view, which is all that is necessary to see all of the planet's features appear on its silhouette without repeating.

\begin{figure}
\centering
\includegraphics[width=8.5cm]{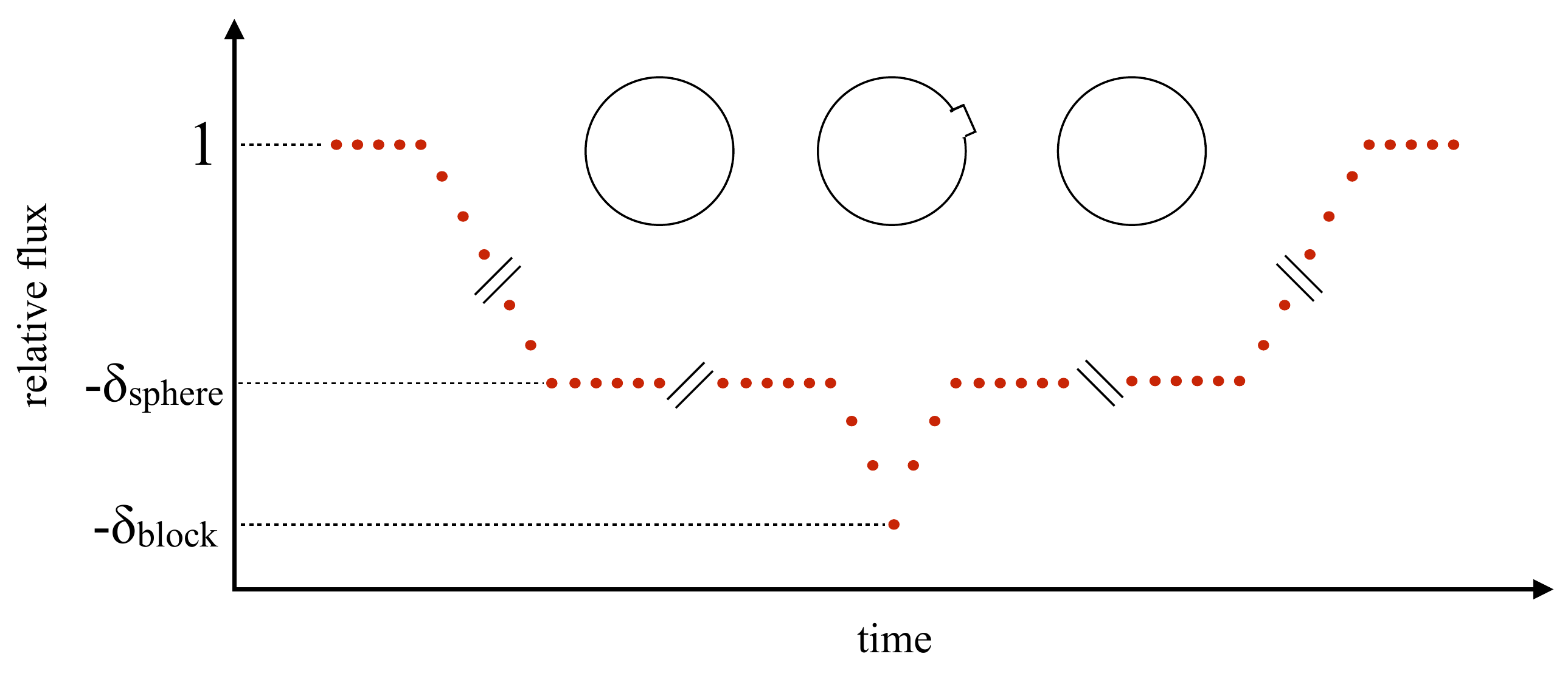}
\caption{Demonstration of how the Himalayan block would affect the bottom of the transit light curve as it appears on the silhouette and moves below the horizon when the planet rotates. In the silhouettes above the plot, the planet's radius is reduced to 1\% of its original size, but the Himalayan block is unchanged. We do this because the block would be otherwise unnoticeable to the human eye. The transit depths are calculated with the original planet radius.}
\label{fig:block}
\end{figure}

As our hypothetical planet rotates and the Himalayan block moves into and out of view, the change in silhouette will result in different transit depths, as can be seen in Fig.~\ref{fig:block}.

The scatter produced by the Himalayan block is $\sim 10^{-8}$ for a 1\,$R_{\odot}$ G-dwarf, $\sim 10^{-6}$ for a 0.1\,$R_{\odot}$ late M-dwarf, and $\sim10^{-4}=100$\,ppm for a 0.01\,$R_{\odot}$ white dwarf. These values are consistent with the amplitudes we estimate in Section 3, where we employ a more physically detailed model.

\newpage
\section{Topographical Model}
\label{sec:model}
\subsection{Overview}

\begin{figure}
\centering
\includegraphics[width=8.5 cm]{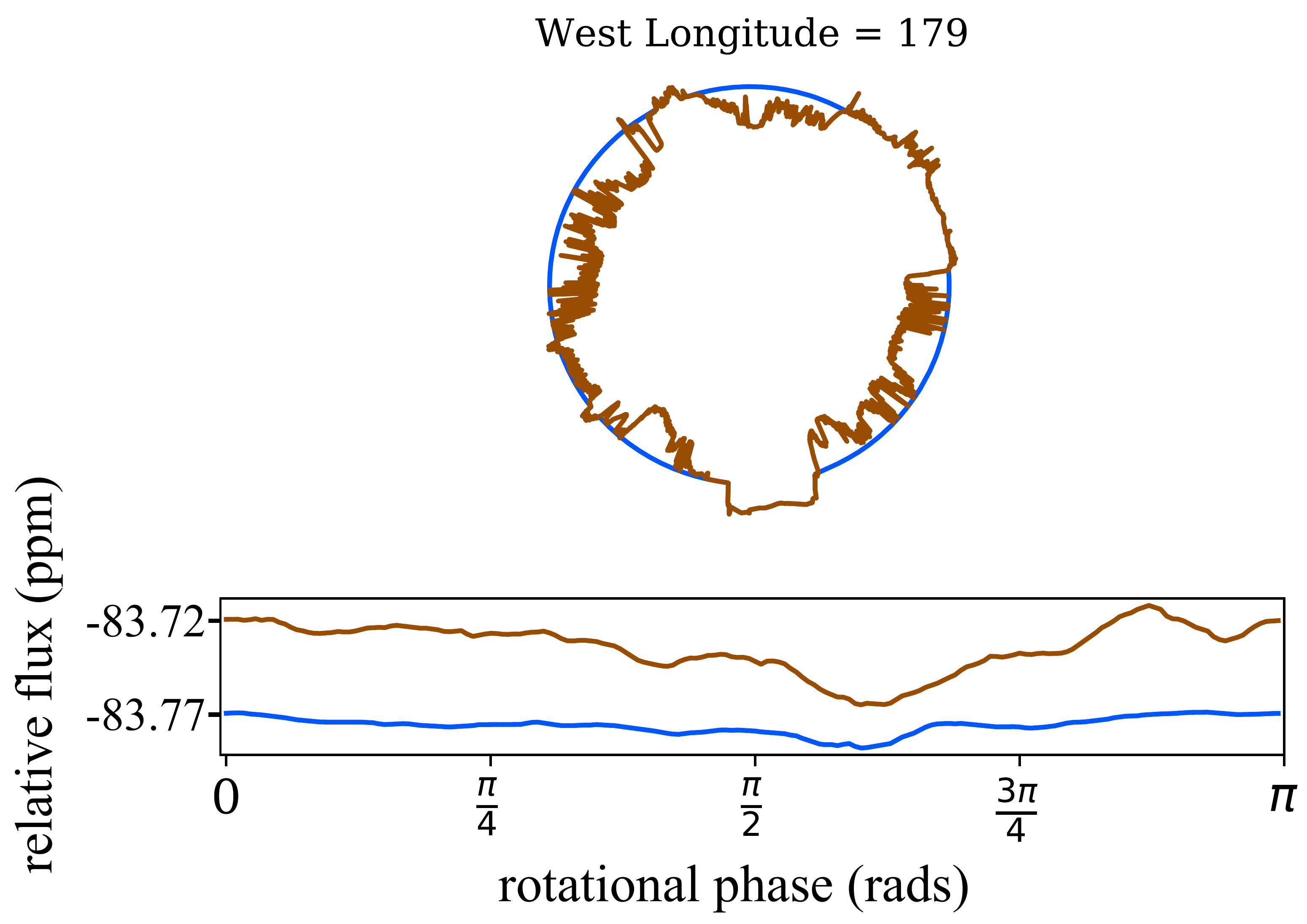}
\caption{Exaggerated silhouette for Earth at a single rotational phase. Here the radius of Earth has been reduced to 1\% its actual value to make the topographical features stand out. The graph at the bottom shows the actual transit bottom for Earth orbiting a sun-like star without noise. The blue line is the light curve of Earth with oceans, and the brown line is what the light curve would look like if oceans were removed. A video of Earth rotating and the resulting transit bottom can be found here: \href{https://github.com/momctier/exo/blob/master/sil.gif}{https://github.com/momctier/exo/blob/master/sil.gif}}
\label{circ}
\end{figure}

Whether from internal processes like movement of tectonic plates or external ones like asteroid bombardment, real planets have more topographical features -- or bumps -- than a single Himalayan block (see Figure~\ref{circ}), which complicates the calculation of bumpiness and scatter. To get a physically motivated relationship between a planet's bumpiness $B$ and its transit depth scatter $\sigma_B$, we use elevation data for the rocky bodies in our solar system: Mercury \citep{becker:2016}, Venus \citep{ford:1992}, Earth, Mars \citep{smith:2001}, and the Moon \citep{smith:2010}. 

In this section, we describe how we first use that data to find $B$ and $\sigma_B$ for the bodies in our solar system. We then use those values to derive a general relationship between bumpiness and transit depth scatter, which we provide in the form of an equation that uses $\sigma_B$ and $R_*$ as inputs and returns $B$. 

\subsection{Elevation Data}
We queried elevation data for the Earth from the Shuttle Radar Topographic Mission (SRTM) database. SRTM provides geodetic data, which are elevations with respect to Earth's reference ellipsoid, an idealized equipotential surface. The database also provides orthometric data, which are elevations with respect to Earth's geoid, or the mean ocean surface. With these two types of data, we can calculate the cross sectional area of Earth both with and without oceans.\\\\ 

\begin{figure}
\centering
\includegraphics[width=8.5 cm]{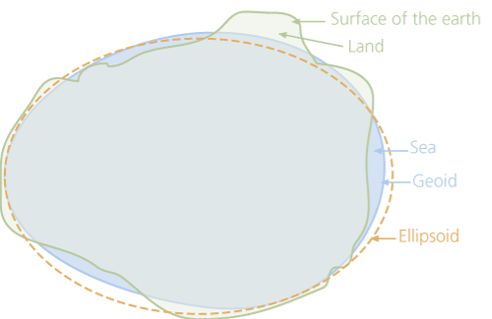}
\caption{Illustration of reference ellipsoid, geoid, and topographical surface \citep{fraczek:2003}}
\label{land}
\end{figure}

\begin{equation}\label{on}
R_{\mathrm{wet}}=H+G_{\mathrm{mod}}+u
\end{equation}

\begin{equation}\label{off}
R_{\mathrm{dry}}=H+E
\end{equation}

Here H is the distance from the center of the planet to the reference ellipsoid, G is the height with respect to the geoid, E is the height with respect to the ellipsoid, and u is the difference between G and E known as undulation. $G_{\mathrm{mod}}$ is the orthometric elevation modified in such a way that any negative numbers are converted to 0 so that features under the ocean surface are ignored.

Elevation data for Mercury, Venus, Mars, and the Moon were publicly available on the U.S. Geological Survey (USGS) website. 

We also received an additional set of Mars elevation data from the Goddard Institute for Space Study (GISS) with different resolution than the USGS data, which allowed us to test the effect that different resolutions had on the calculated bumpiness value.

\begin{table}
\centering 
\begin{tabular}{ccccc} 
\hline\hline
World & Source & Lat. Res. & Long. Res. & $\alpha^{\mathrm{a}}$ \\ [0.5ex] 
\hline
Mercury	& USGS				& $0.080^{\circ}$		& $0.080^{\circ}$		& $5^{\circ}$ \\
Venus		& USGS				& $0.044^{\circ}$		& $0.044^{\circ}$		& $5^{\circ}$ \\
Earth		& SRTM				& $0.02^{\circ}$		& $0.02^{\circ}$		& $3.2^{\circ}$ \\
Moon		& USGS				& $0.0975^{\circ}$	& $0.0975^{\circ}$	& $5^{\circ}$ \\
Mars		& USGS				& $0.078^{\circ}$		& $0.078^{\circ}$		& $6.7^{\circ}$ \\
Mars		& GISS				& $2^{\circ}$				& $2.5^{\circ}$			& $6.7^{\circ}$ \\ [1ex]
\hline 
\end{tabular}
\caption{
Table showing the source of data and data resolution for each rocky body. $^a$ The $\alpha$ angle is explained in Section~\ref{proj} and is calculated for Earth and Mars, but estimated for the other bodies because their highest points weren't known before analyzing the data.
}
\label{tab:datatable} 
\end{table}

\subsection{Defining ``Bumpiness''}
\label{bump}

We searched for an existing term that would describe the variance in elevation of a rocky body. In Earth science fields, there is a term called ``relief," which refers to the difference between the highest and lowest points in a given region. This term didn't suit our specific needs, so we created a new term to quantify the presence of topographical features on a rocky body. We call that term ``bumpiness."

When choosing how to define bumpiness of the rocky bodies in our solar system for use in our topographical model, we had three requirements in mind. 

First, bumpiness should be an inherent characteristic of the planet. By that, we mean a planet's bumpiness shouldn't change based on our observation of it. For example, viewing angle and host star radius could both affect the way topography influences a planet's light curve, but neither should affect a planet's bumpiness. 

Second, the definition should encode the planet's radius. An Everest-sized mountain on an otherwise featureless Mercury provides more contrast to the average planet radius than an Everest on an otherwise featureless Earth, and should result in a higher bumpiness value.

Third, the definition should be an assessment of global average features. By that, we mean that it should not just take into account the largest feature on the planet, but add up the individual contributions from all the planet's features. 

We considered many definitions for bumpiness, but ultimately the only one we could think of that met all three criteria was to take the standard deviation of the radial distances from the center of a planet to every point on the planet's surface.

The data we obtained from USGS ($0.078^{\circ} \times 0.078^{\circ}$) yields a bumpiness value of $\approx 2977$ meters. The data we obtained from GISS ($2^{\circ} \times 2.5^{\circ}$) yields a bumpiness value of $\approx 2948$ meters. The values are extremely close given the orders of magnitude difference in their resolutions. We can’t say whether our resolutions yield under- or overestimates for bumpiness, but our results are likely extremely close to the truth.

\subsection{Projected Geometry}\label{proj}
To simulate a planet's light curve and find its $\sigma_B$, we need to calculate the cross-sectional area of the planet at every rotational phase, which is advanced by stepping through longitude in $1^{\circ}$ increments, so chosen to balance precision and computational convenience.

At every rotational phase, a cross-section - or ``great circle" - is defined by a west longitude and an east longitude $180^{\circ}$ to the east. The first step in calculating the cross-sectional area is finding the distance from the center of the planet to every point on the projected silhouette. This distance might not necessarily be the elevation at the point on the great circle because the silhouette could be dominated by a taller point that sits on a longitude that isn't used to define the great circle. In other words, a taller point might be behind or in front of the point on the great circle from our point of view. 

To account for this projected geometry issue, we first define an ``$\alpha$ angle," which is the number of degrees a planet would have to rotate in order for its tallest feature to disappear below the horizon after it appears on the great circle. For example, Everest is the tallest mountain on Earth, and Earth would have to rotate $3.2^{\circ}$ from the time when Everest is on the great circle before the tip of Everest would no longer appear in the planet's silhouette. 

\begin{equation}\label{alpha}
\alpha = \cos^{-1}\left ( \frac{R\: \mathrm{sin}(90-l)}{R\: \mathrm{sin}(90-l)+h} \right )
\end{equation}

Here $l$ is the latitude of the highest point on the planet and $h$ is the elevation of the highest point. 

For both the west and east longitude lines used to define the great circle, we step through in latitude, and at every latitude, we find the maximum projected height. The projected height of point A is  

\begin{equation}\label{height}
D = d\:\cos\left ( \beta \right )
\end{equation}

where $d$ is the actual elevation of point A and $\beta$ is the angle between A and the point along the great circle. $\beta$ ranges from $-\alpha$ to $+\alpha$. 

The projected heights give us the distances from the center of the planet to every point along the silhouette. The area of the silhouette can then be found using trapezoidal integration.

\begin{equation}\label{area}
A_P = \sum_{i}^{N}R_i^2 \: \tan\left ( \frac{\mathrm{lat\: res}}{2} \right )
\end{equation}

Once the planet's cross-sectional area is found, the transit depth can be calculated using Equation~\ref{depth}.

\subsection{Defining Scatter}

The projected geometry method described above provides a list of transit depths for every rotational phase, so $\sigma_B$ is easily defined as the standard deviation (std) of all of the depths. We use these $\sigma_B$ values to derive the general relationship between $B$ an $\sigma_B$ in Section~\ref{rel}.

\begin{figure*}
\centering
\includegraphics[width=\textwidth]{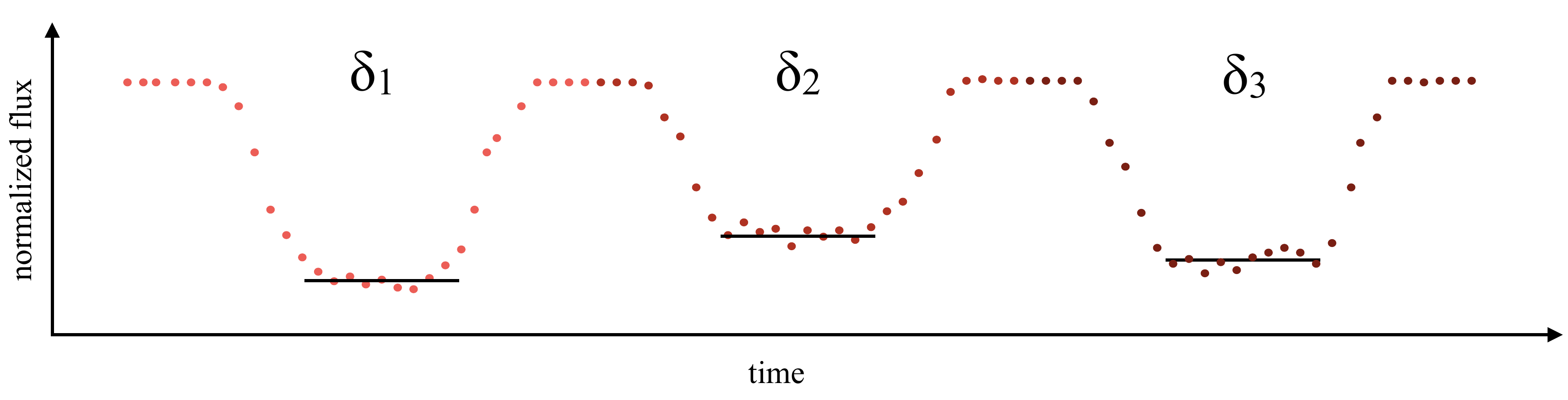}
\caption{Individual transits shown with varying average depths. Here, $\delta_1 < \delta_3 < \delta_2$ and the scatter in average depths can be calculated to find the bumpiness of the transiting planet.}
\label{series}
\end{figure*}

\begin{figure}
\centering
\includegraphics[width=8.4cm]{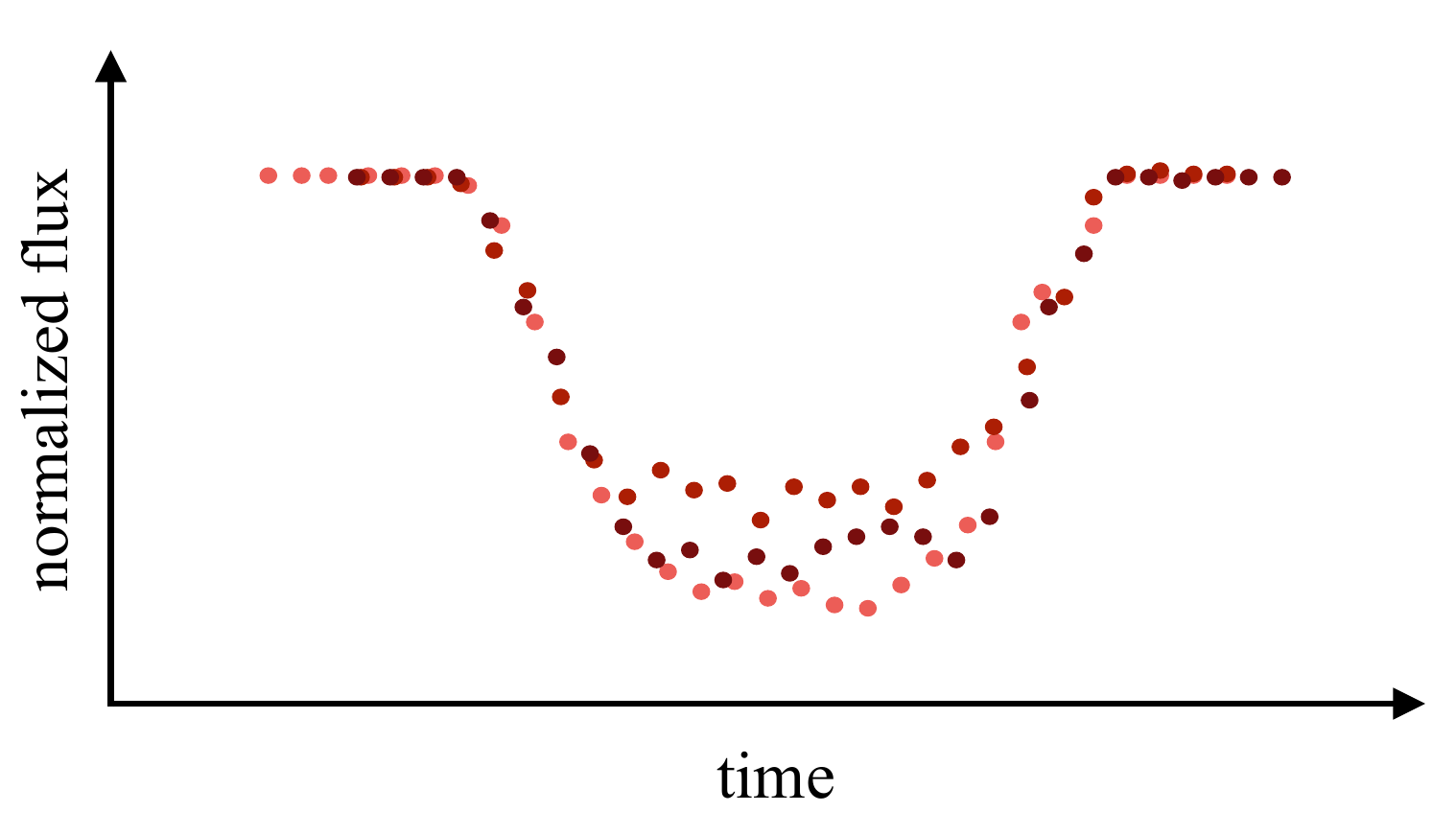}
\caption{Folded lightcurve of the same data shown in Figure~\ref{series}. The scatter calculated from this data is more accurate and should be used to find the bumpiness of the transiting planet.}
\label{folded}
\end{figure}

In the Earth's case, a transit would last for up to approximately 13 hours - more than half a rotational period - and thus we would observe the full range of phases necessary to calculate $\sigma_B$. However, other transiting planets may not last long enough for us to observe a sufficient fraction of a planet's rotation. Instead, individual transits will show various ranges of rotational phases, which may or may not overlap depending on the planet's rotational period. Unfortunately, this means that we might not observe every rotational phase. When this is the case, the $\sigma_B$ we find in our observations may be less accurate than the ``true" value. In this section, we merely describe the two ways to evaluate $\sigma_B$.  

The first way is to find the average depth of each individual transit (see Figure~\ref{series}), and take the std of those averaged depths. The second way is to fold the light curve and take the std of the folded transit depth (see Figure~\ref{folded}). 

The second method provides a more accurate $\sigma_B$ value because the process of taking the average depths in the first method smooths out the extreme points in the transit bottoms. Those extreme points are caused by extreme features, such as mountain ranges and trenches, which is what we're hoping to find. Taking the std of the folded light curve exploits the full data set and is ultimately limited by the observing cadence, which can be generally assumed to be much less than the rotational period.

\newpage
\subsection{Scatter-Bumpiness Relation}
\label{rel}

After finding $B$ and $\sigma_B$ for Mercury, Venus, Earth, Mars, and the Moon, we used those values to find a general relationship between bumpiness and transit depth scatter. This relationship provides a means to convert from an observed excess scatter, $\sigma_B$, to a geophysically-motivated measure of elevation variance, $B$, and thus place the observed body in a broader context.

\begin{figure}
\centering
\includegraphics[width=8.4cm]{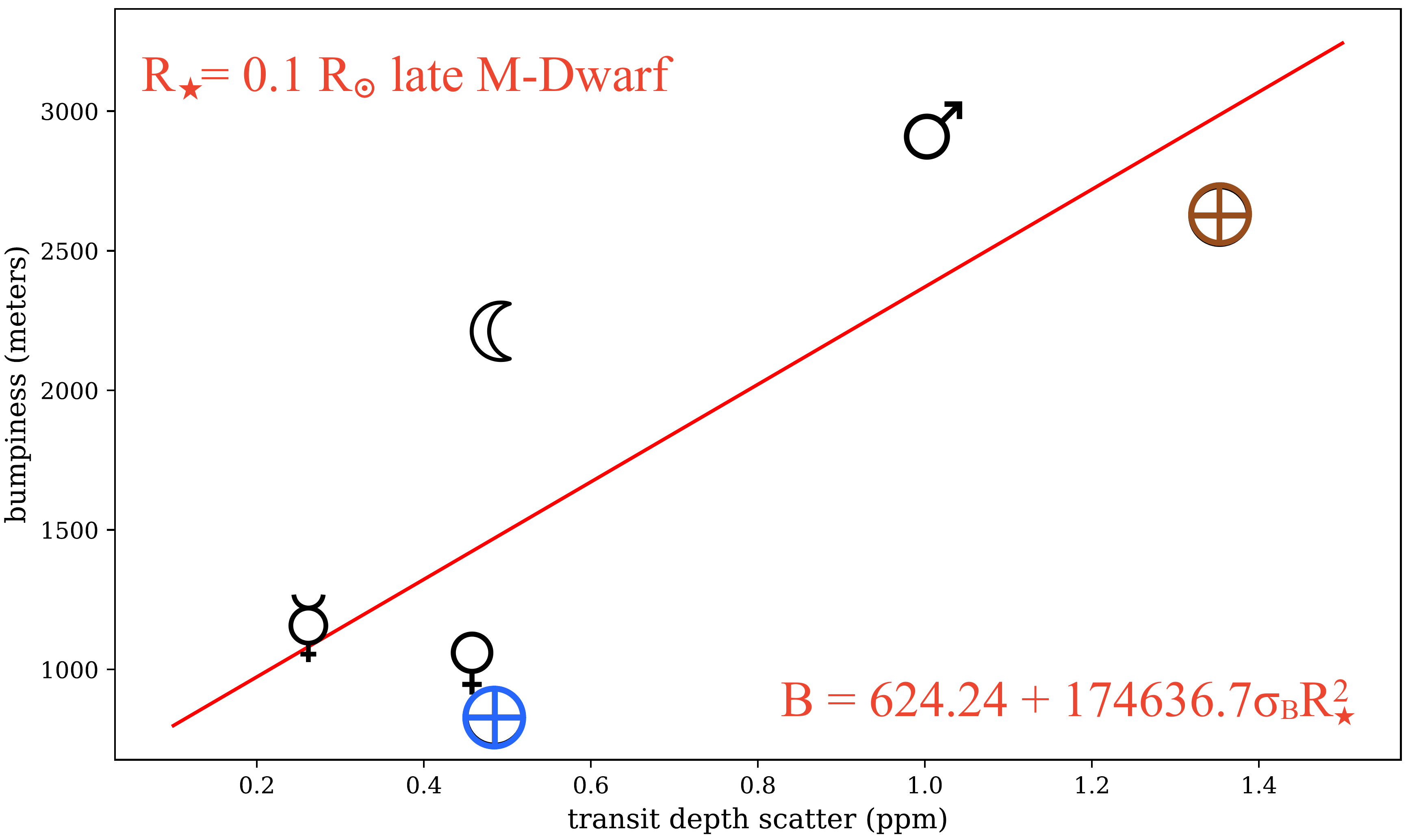}
\caption{Relationship between $B$ and $\sigma_B$ for a 0.1\,$R_{\odot}$ late M-dwarf. The blue earth symbol represents a wet Earth and the brown symbol represents a dry Earth, as in Figure~\ref{circ}.}
\label{fig:rel}
\end{figure}

In Figure~\ref{fig:rel}, we show the relationship between $B$ and $\sigma_B$ for a 0.1\,$R_{\odot}$ late M-dwarf, which can be calculated using the following equation:

\begin{equation}\label{bump_eqn}
B = 624.24 + 174636.7\:\sigma_B\: R_*^2,
\end{equation}

which will return a bumpiness value in meters if $\sigma_B$ is in parts per million and $R_*$ is in solar radii.

Because of how we define bumpiness, $B$ is constant for each planet, but $\sigma_B$ changes because a light curve's depth -- and therefore the scatter in that depth -- depends on the radius of the host star. This is why there's a term for $R_*$ in our equation.

We recognize that this is not a one-to-one deterministic relationship, and thus the conversion itself will introduce some extra uncertainty. Improving the correspondence of the relation and/or adding a formal probabilistic component to it would be useful topics to investigate in future work, but we proceed with this relation as a first estimate.

\section{Feasibility}
\label{sec:feasibility}

\subsection{Necessary Photometric Precision}

Thus far, we have ignored the unavoidable existence of photometric noise, and so here we estimate the detectability of the predicted excess scatter from bumpiness, $\sigma_B$, in the presence of a stochastic noise process described by a variance $\sigma_n^2$. 

Consider a photometric time series of regular cadence where the covariance of the time series due to instrument plus photon noise may be represented by a diagonal, multinormal distribution with homoscedastic variances, $\sigma_n^2$.

For data occurring outside of the transits, the expectation value for the standard deviation of the time series photometry is simply $\mathrm{E}[\sigma_{\mathrm{obs}}]=\sigma_n$. However, each random segment of $N$ data points will exhibit slightly different $\sigma_{\mathrm{obs}}$ values, due to the fundamental stochastic nature of the noise-generating process. Specifically, in the limit of a large number of points, the central limit theorem may be used to show that $\sigma_{\mathrm{obs}}$ will be normally distributed and described by $\sigma_{\mathrm{obs}} \sim \mathcal{N}[\sigma_n,\Upsilon_n]$, where

\begin{equation}
\Upsilon_n = \frac{\sigma_n}{\sqrt{2N}},
\end{equation}

and $N$ is the number of data points falling within each segment.

For the in-transit data, topographic variations lead to increased photometric scatter, as described earlier, modifying the expectation value for the observed scatter to $\mathrm{E}[\sigma_{\mathrm{obs}}] = \sqrt{\sigma_n^2+\sigma_B^2}$. This may be compared to the range expected due to random variations, as described above, in order to estimate the one-sided $p$-value significance of a particular choice of $\sigma_B$. Whilst we do not recommend the use of $p$-values for model selection, it is suitable for the estimation of sensitivity, as required here. The significance level, $f$, of the observed scatter being, in part, the result of bumpiness may therefore be estimated as 

\begin{align}
f &= \frac{ \sigma_{\mathrm{obs}} - \sigma_n }{ \Upsilon_n },\nonumber\\
\qquad&= \frac{ \sqrt{\sigma_n^2+\sigma_B^2} - \sigma_n }{ \frac{\sigma_n}{\sqrt{2N}} }.
\end{align}

The above may be re-arranged depending on the desired purpose. For example, for a fixed $\sigma_B$, $\sigma_n$ and desired $f$, the number of data points, $N$, to achieve a detection would be given by

\begin{align}
N &= \frac{(f^2/2)}{(\sqrt{1+(\sigma_B/\sigma_n)^2} - 1)^2}.
\end{align}

\subsection{Estimated Detectability for a Non-Tidally Locked Planet}

Due to a combination of its small size, low surface gravity, and active internal volcanism, Mars is the bumpiest body in the inner Solar System (see Figure~\ref{fig:rel}). As an optimal case for detecting bumpiness, we consider here a white dwarf parent star with an orbiting Mars-sized planet. We place the planet at 0.01\,AU (around the center of the long-lived habitable-zone \citealt{agol:2011}), which for a typical white dwarf mass of $0.6$\,$M_{\odot}$ leads to an orbital period of 11.3\,hours and a flat-bottomed transit duration of $T_{23}=27$\,seconds (for an edge-on impact parameter).

Consider observing a large number of transits around a target star where the planet appears in a random phase during each transit, which means that the planet is not tidally locked to the star. If the rotation period is much longer than the transit duration of around a minute, then the detectability may be estimated by considering the variations in transit depths occurring between each epoch. From our earlier simulations, we estimate Mars would introduce variations of $\sigma_B = 119.4$\,ppm between each transit depth, which would average a 28.2\% decrease in flux (assuming $R_{\star} = R_{\oplus}$).

Consider monitoring these transits with one of the extremely large telescopes currently under development, in the region of 30-80\,m in diameter. To estimate photometric sensitivity, we scale from a repeatable example of precise ground-based transit photometry coming from the series of Southworth papers on extreme telescope defocussing. On the 1.54\,m Danish telescope in R band, these results are able to obtain a precision of $\sim 0.5$\,mmag for a 2\,minute exposure for an $R\sim12$ target, e.g. WASP-5 \citep{southworth:I} and WASP-4 \citep{southworth:II}. Scaling to a fiducial 1\,meter telescope, we estimate that a ground-based precision of $\sigma_0=1.2$\,mmag per minute on an $R=12$ star is quite reasonable.

We estimate the corresponding R magnitude of our white dwarf (WD) again by a simple scaling argument. The transit probability of our hypothetical WD planet would be $0.65$\% and thus even if the occurrence rate of such objects were 100\%, one should expect to survey $\sim150$ WDs before finding such an example. Given the local density of WDs of $4.7 \times 10^{-3}$\,pc$^{-3}$ \citep{holberg:2002}, then this would imply a distance of $31.9$\,pc, or $68.8$\,pc if the occurrence rate were lower at 10\%. The R band magnitude of such a star can be estimated by scaling from a well-known 0.6\,$M_{\odot}$ WD, L 97-12, which is $R=13.58$ at a distance of $d=7.92$ pc. We therefore estimate our hypothetical WD would have an $R$ magnitude of $16.6$ for a 100\% planet occurrence rate, or $R=18.3$ for a 10\% occurrence rate.

Using our earlier result, we may now estimate the photometric precision obtained for a single transit of a telescope of diameter $d$ to be

\begin{align}
\sigma_n &= (1200\,\mathrm{ppm}) \Big(\frac{1\,\mathrm{meter}}{d}\Big) 
\Big( \frac{ 60\,\mathrm{secs} }{ T_{23} } \Big)^{1/2} \sqrt{10^{ 0.4 (R-12)}}.
\end{align}

\begin{figure}
\centering
\includegraphics[width=8cm]{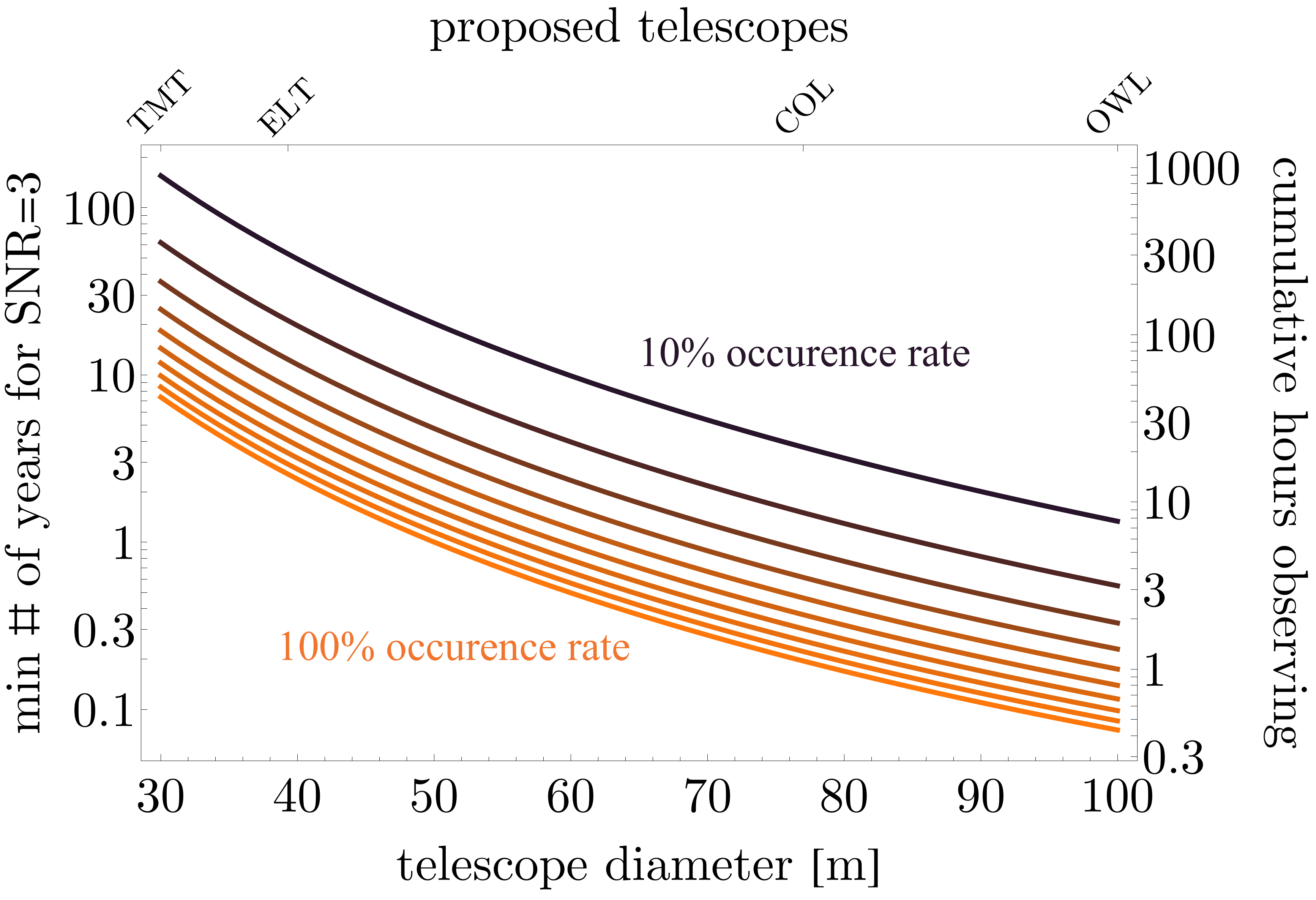}
\caption{Plot showing how much time telescopes of different sizes would have to observe to make a $3\sigma$ detection of bumpiness on a planet orbiting a white dwarf. The right y-axis shows the total number of hours observed, while the left y-axis shows the corresponding number of years of wall time.}
\label{fig:sens}
\end{figure}

Combining our estimates for scatter caused by bumpiness with the photometric noise estimates described above, we find that super telescopes could indeed plausibly detect topography. For example, as shown in Figure ~\ref{fig:sens}, Colossus could detect topography to $3\sigma$ confidence for a 10\% WD planet occurrence rate using $\approx400$ transit measurements. Though this sounds large, each transit lasts just a few minutes, which means that the actual observing time would be on the order of 20 hours. For the smaller ELT configuration and a 100\% occurrence rate, detections could still plausibly be achieved with the same amount of observing time.

\begin{figure*}
\centering
\includegraphics[width=\textwidth]{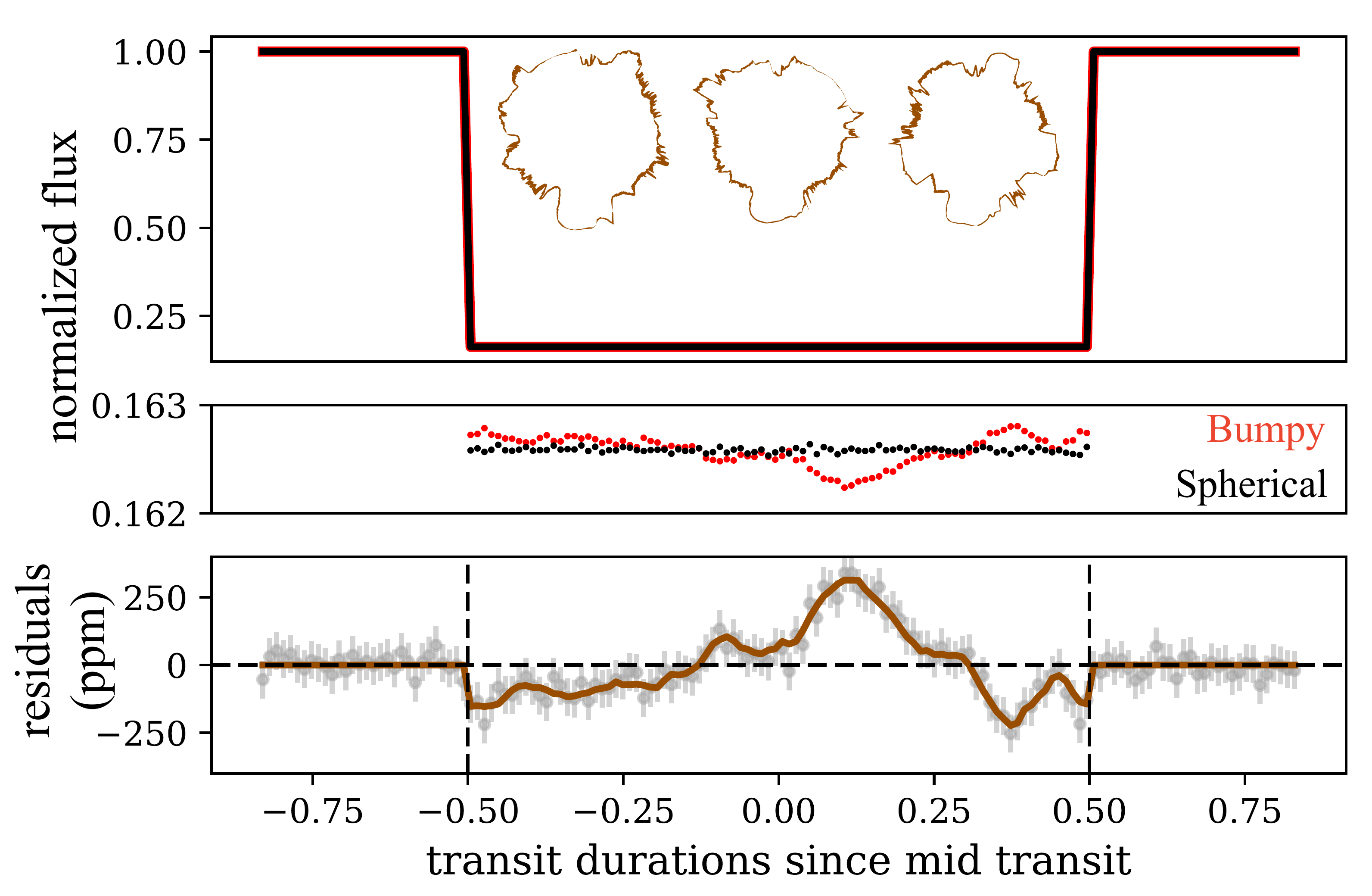}
\caption{\textbf{Top}: Transits of a dry Earth with features (in red) and an idealized spherical Earth (in black) in front of a $.01R_{\odot}$ white dwarf with noise of 20 ppm added ($20\sigma$ detection). The exaggerated silhouettes of Earth at different rotational phases are shown in brown. \textbf{Middle}: Zoomed-in frame of the bottom of the light curve in the top panel. \textbf{Bottom}: Residual plot showing the difference between the realistic and idealized transits. Grey shadows show the error bars on the residuals equal to 50\,ppm. Dashed lines are to illustrate that residuals deviate from 0ppm only inside the transit.}
\label{resid}
\end{figure*}

\subsection{Sources of False Positive}
Aside from issues of sensitivity and precision, there are some phenomena that would produce extra scatter in the light curve and appear to increase $\sigma_B$.  

\begin{itemize}
  \item \textbf{Stellar Activity:} Asteroseismic pulsations would add to the light curve scatter, but the effect should be present both in and out of transit, so it can be measured and subtracted from the data. Stellar flares would produce spikes in the light curve orders of magnitude too extreme to be caused by topographical features, and could be removed as outliers. White dwarf atmospheres are theoretically too hot to have convective star spots \citep{brinkworth:2007}, but if any are ever discovered, they'll be chromatic, so they can be characterized using multi-wavelength observations.
  
  The periodic variations seen in WDs from Kepler are 60 to 2000ppm \citep{maoz:2015} and the intrinsic variations we predict in this work for the case of a Mars transiting a WD are $\sim120$ppm, so certainly the lower end of that range is not going to exclude our ability to detect these features. We stress that the scatter from bumpiness only occurs in-transit, so essentially one would observe increased noise levels directly during the transit events, which could be easily compared to the typical behavior out of transit. The more challenging case is when the intrinsic stellar noise exceeds that of the bumpiness. However, given that these intrinsic variations are periodic, then it would appear quite reasonable that such intrinsic variations could be cleaned using techniques through the use of frequency filters. We therefore remain optimistic that exotopography could be distinguished.
  \item \textbf{Exomoons:} Quickly orbiting moons that move in and out of the silhouette as a planet transits would cause variation in the transit depth. \citet{szabo:2011} identified this as a way to detect exomoons directly from light curves. That effect can be distinguished from $\sigma_B$ because scatter from topographical features only appears in in-transit data, while scatter from exomoons also appears slightly outside of the ingress and egress points. Figure~\ref{resid} illustrates the effect that topographical features can have on a light curve by showing an oceanless Earth transiting a white dwarf, both with surface features (red) and without (black). The residual plot in the bottom panel shows that the effect only appears within the transit.
  
\end{itemize}

\subsection{Sources of False Negative} \label{neg}
There are also many phenomena that could decrease or eliminate $\sigma_B$.

\begin{itemize}
\item \textbf{Clouds:} On Earth, water vapor clouds tend to congregate around mountains for several reasons, one of which is that the presence of the mountain forces air to move up into colder regions of the atmosphere where water vapor can condense into clouds \citep{houze:2011}. There is no reason to expect that water vapor clouds on exoplanets wouldn't follow the same pattern. Clouds absorb most Far-infrared light, which is the region of the electromagnetic spectrum for which the James Webb Space Telescope has been optimized \citep{beichman:2014}. Infrared atmospheric windows could provide an opportunity to see through the clouds, but they vary with a planet's environmental conditions, such as water vapor content. This means that ever-present, thick clouds could obscure our view of a planet's surface, and yield little or no $\sigma_B$, however there are likely many planets, like Mars, that have either no atmosphere or little cloud coverage.

\item \textbf{Obliquity:} In our work, we assume a planet obliquity of 0$^{\circ}$ (axis of rotation perpendicular to the line of sight), which causes the maximum observable transit depth scatter from bumpiness. The opposite extreme is an obliquity of 90$^{\circ}$ (axis of rotation pointing along the line of sight), which would yield no $\sigma_B$ because the same features would always appear on the silhouette. Planets with a 90$^{\circ}$ obliquity are rare, though, with Uranus being the only such planet in our solar system. 

\item \textbf{Oceans:} Oceans can obscure many topographical features and make a planet appear less bumpy than it really is, however this could actually be used to identify ocean-rich worlds if other sources of false negative can be ruled out.

\item \textbf{Depressions:} Valleys, trenches, and chasms are more difficult to observe than mountains because the higher surface features on either side would likely dominate a silhouette. Any depressions that do appear in the silhouette are rather large and run perpendicular to the axis of rotation.
\end{itemize}

\subsection{Tidal Locking}\label{tidal}

Tidal locking is neither a false negative nor a false positive, but it does present challenges to our exotopography method. Tidal locking obscures information about a planet's surface by limiting the range of rotational phases we can observe. In order for our method to work, we need to either observe a planet at different rotational phases in individual transits or observe a planet rotating as it transits its host star.  




The ideal case we've presented in this work for detecting scatter from bumpiness is that of a rocky planet orbiting a white dwarf with a short period ($\sim10$ hours). While tidal locking is certainly a possibility for such planets, we think that in general one cannot assume they will suffer this fate.

What is most important to consider is whether the planet is rotationally synchronized to the star, not necessarily whether it is tidally locked. For example, an eccentricity in the orbit exceeding $\sim0.1$ may result in supersynchronous rotation (see \citet{goldreich:1966}; \citet{barnes:2008}; \citet{ferraz:2008}; \citet{correia:2008}), meaning that their rotation rate is fixed by tidal torques, yet they do not rotate synchronously. It is quite possible that such planets would get caught in spin-orbit resonances such as 3:2 (e.g. Mercury \citep{noyelles:2014}) or 2:1 \citep{rodriguez:2012}, and that would provide a chopping nature to the bumpiness effect described in this work. Nevertheless, we can expect the overall amplitude to be comparable to our earlier predictions. Even in the pessimistic case of a 1:1 spin-orbit resonance, one should still expect to see libration of the planet's phase, similar to that of the Moon, for even small eccentricities. Again, these librations will yield a topographical bumpiness effect of similar amplitude to that predicted here.

A second mechanism by which planets might be reasonably expected to avoid tidal locking is via thermal tides induced on their atmospheres. \citet{leconte:2015} showed how even thin atmospheres can prevent tidal locking for habitable-zone planets of low-mass main-sequence stars. To our knowledge, there are no studies or clear scaling as to how this effect will vary for white dwarf planets, as the results of \citet{leconte:2015} are based on detailed GCMs. We think that these arguments demonstrate that a rotationally synchronized planet is not an inevitable outcome and thus exotopography remains a plausibly detectable signal under the conditions we consider in this work.

\section{Discussion}
\label{sec:discussion}
In this work, we've investigated the detectability of exotopography by introducing a novel method to infer mountainous features through the study of transit light curves. At its root, this method is based on a simple concept: mountains, trenches, and other surface features appear in a planet's silhouette as it transits its host star. As the planet rotates, its changing silhouette will yield observable scatter in transit depth of the light curve.

We've shown that, even with the technology that will become available in the coming decades (ELT, OWL, and Colossus), it will be near impossible to detect bumpiness on planets orbiting Sun-like stars and M-dwarfs due to astrophysical and instrumental noise. It will be feasible, however, to achieve the precision necessary to observe this scatter for rocky planets orbiting white dwarfs. But why should we even care about finding exotopography? 

Finding the first evidence of mountains on planets outside our solar system would be exciting in its own right, but we can also infer planet characteristics from the presence and distribution of surface features.

For example, a detection of bumpiness could lead to constraints on a planet's internal processes. Mountain ranges like the Himalayas on Earth form from the movement and collision of tectonic plates \citep{allen:2008}. Large volcanoes like Olympic Mons on Mars form from the uninterrupted buildup of lava from internal heating sources. A high-bumpiness planet is likely to have such internal processes, with the highest bumpiness values resulting from a combination of low surface gravity, volcanism, and a lack of tectonic plate movement. Truly low-bumpiness planets are less likely to have these internal processes. On such planets, surface features are likely caused by external factors like asteroid bombardment. 

Other planet characteristics could be derived using a combination of this exotopography method and future work. For example, we mention in Section~\ref{neg} that oceans will obscure topographical features from view, making a planet appear less bumpy than it really is. Future work can be done to distinguish between a truly low-bumpiness planet and a planet with an ocean.

We also encourage future work that explores the potential to use exotopography to constrain a planet's rotational period. Our method for finding a planet's bumpiness relies on seeing the planet rotate, but depending on the relationship between a planet's rotational and orbital periods, we might not observe a sufficient fraction of rotational phases. If we could constrain the planet's rotational period, we could discern the range of rotational phases observed, which would reveal the accuracy of a calculated bumpiness value. 

To apply our method to exoplanet observations, we will need to find significance in what we have considered noise for decades. This means that we will have to consider innovative ways to reduce noise and remove outliers so that we don't accidentally eliminate useful information from the light curve. 

We encourage members of the community to use our exotopography method, as well as any future improvements and expansions, to aid in our mission of adding texture to worlds outside our own. 

\section*{Acknowledgements}

This research has made use of data provided by Michael Way at GISS and code provided by Trent Hare at USGS. 

We thank members of the Cool Worlds Lab for stimulating conversations on this topic.







\bsp	
\label{lastpage}
\end{document}